\documentclass[aps,preprint,floats,nofootinbib]{revtex4}
\usepackage{graphicx}
\usepackage{epsf}
\usepackage{feynarts}
\def\lsim{\mathrel{\raise.3ex\hbox{$<$\kern-.75em\lower1ex\hbox{$\sim$}}}}
\def\gsim{\mathrel{\raise.3ex\hbox{$>$\kern-.75em\lower1ex\hbox{$\sim$}}}}

\begin{document}
\hfill$\vcenter{\hbox{\large OUTP-04/23P}}$

\vskip 0.5cm

\title {Gauge Mediated Supersymmetry Breaking and Multi-TeV Gamma-Rays from the Galactic Center} 
\author{Dan Hooper$^1$ and John March-Russell$^2$} 
\address{$^1$ Astrophysics Dept., University of Oxford, Denys Wilkinson Building, Oxford OX1 3RH, UK;\\
$^2$ Rudolf Peierls Centre for Theoretical Physics, University of Oxford, 1 Keble Road, Oxford OX1 3NP, UK}
\date{\today} 

\bigskip
\bigskip

\begin{abstract}

Recently, the HESS telescope has observed a gamma-ray spectrum from
the galactic center extending to at least $\sim$10 TeV. Although it
has been suggested that this could be the product of annihilating dark
matter particles, the candidates most frequently discussed (such as neutralinos) are
far too light to account for this flux. In this letter, we consider
stable particles from the messenger sector of gauge mediated
supersymmetry breaking models as an alternative dark matter
candidate.  We find that a 20 to 30 TeV messenger state can provide a thermal relic density consistent with the measured dark matter density of the universe and can indeed generate the spectrum observed by HESS.

\end{abstract}

\maketitle 

\newpage

\section{Introduction}

Recently, the High-Energy Spectroscopic System (HESS) collaboration has reported the detection of gamma-rays from the region of the galactic center with energies ranging between 
approximately 250 GeV and 10 TeV  \cite{hess}. Two other Atmospheric Cerenkov Telescopes 
(ACTs), Cangaroo-II~\cite{cangaroo} and Whipple~\cite{whipple} have also reported positive detections of very high-energy gamma-rays 
from the region, although with considerably lower significance.\footnote{The spectrum reported by the Cangaroo collaboration differs 
substantially from the HESS results. The results of the Whipple telescope, which are limited to an integrated flux, are only 
marginally consistent with HESS.}

Currently, the origin of this very high-energy emission is unknown. Efforts have been made to explain the data with astrophysical 
processes involving the supermassive black hole known to exist at the galactic center \cite{Aharonian:2004jr,dermer}. Although these 
are certainly possible explanations, the source of the observed emission is not yet clear.

Gamma-ray observations of the galactic center have long been studied as one of the most promising methods to search for particle 
dark matter \cite{gammaraysdark}. The most popular dark matter candidates~\cite{dmreview}, such as the lightest neutralino in 
$R$-parity conserving supersymmetric models, are generally much lighter than the highest energy gamma-rays seen by HESS, and 
therefore cannot possibly generate the observed spectrum. It has been shown that a dark matter particle annihilating through typical 
channels ($q \bar{q}, W^+ W^-, Z^0Z^0$) would require a mass between 12 and 29 TeV to generate this spectrum \cite{horns}. In addition to 
the unacceptable levels of fine tuning which would be required, a neutralino in this mass range would generate far more dark matter during 
thermal freeze-out than is observed. For these reasons, we do not consider a very heavy neutralino to be a likely source of the 
flux observed by HESS.

Models of Gauge Mediated Supersymmetry Breaking (GMSB) provide an interesting alternative to neutralino (or other lightest 
supersymmetric particle) dark matter.  Typically in these models, supersymmetry breaking originates in a strongly interacting sector 
(the supersymmetry breaking sector) and is communicated to the superpartners of the Standard Model (SM) states by SM gauge
interactions with particles within a messenger sector \cite{gmsb}.  It is natural for the lightest messenger particle to be stable, 
electrically neutral and have a mass on the order of a few times 10 TeV.  Such a particle, as we will show, can have an 
annihilation cross section which generates a thermal relic density consistent with the measured dark matter density of the universe. 
In this letter, we calculate the gamma-ray spectrum generated in the annihilations of GMSB messenger dark matter and show that this 
particle can be a viable source of the very high-energy gamma-rays observed by HESS.

\section{Messenger Dark Matter in Gauge Mediated Supersymmetry Breaking Models}

GMSB models typically contain a messenger sector, charged under the SM gauge group, which
communicates supersymmetry breaking to the superpartners of SM states by SM gauge interactions.
Such gauge communication of supersymmetry breaking has the great advantage that unobserved flavour-changing-neutral-current, 
$CP$-violating, and rare decay processes arising from the soft-supersymmetry-breaking terms
are automatically absent, in contrast to most other methods of communication. 
Since supersymmetry breaking is communicated by renormalizable gauge interactions, the fundamental scale
of supersymmetry breaking in GMSB models is much lower than for supergravity communication, possibly as low as 50 to 100~TeV
depending on the interactions between the superymmetry breaking and messenger sectors. 
As a consequence, the Lightest Supersymmetric Particle (LSP) is the gravitino, with mass, $m_{3/2}\sim 1-10$ eV.
Such a mass is far too small to contribute substantially to the dark matter density of the universe.
Thus, in this class of models, it is necessary to look beyond the LSP for the constituents of dark matter.
Fortunately in the simplest class of GMSB models, the messenger sector naturally contains a stable state
with mass of a few ten's of TeV.  \footnote{In some GMSB models there are additional stable SM-neutral states
arising from the supersymmetry breaking sector, while in others the messenger and supersymmetry-breaking sectors are
integrated into one sector leading to changes in the phenomenology \cite{gmsb,nonmin}.  We will not discuss these non-minimal
models in this letter.}

Messenger particles in GMSB models have been studied as potential dark matter candidates in the past. In Refs.~\cite{dgp,han1}, 
minimal supersymmetric models with gauge mediated supersymmetry breaking were studied. In these models, the messenger sector 
consists only of particles carrying SU(3)$\times$SU(2)$\times$U(1) charge, the lightest of which will be stable if the supersymmetry 
breaking sector contains only singlets charged under this group. In substantial portions of the parameter space, the lightest 
messenger state is a massive complex scalar with the SM gauge quantum numbers of a left handed neutrino.

This simple model ultimately fails to provide a viable dark matter candidate, however. With only gauge interactions to contribute to 
its annihilation rate in the early universe, it has been shown that such a particle must be lighter than a few TeV to not overclose 
the universe~\cite{dgp,han1}. In addition to being too light to produce the very high-energy gamma-rays observed by HESS, such a 
dark matter candidate is excluded by direct detection searches.

A very simple way to extend the Minimal Supersymmetric Standard Model (MSSM) is to introduce an additional Higgs singlet. This 
model, the Next-to-Minimal Supersymmetric Standard Model (NMSSM), has several advantages compared to the MSSM. Most notably, it 
provides a natural mechanism for the $\mu$-parameter to obtain a value near the electroweak scale (which is required for acceptable 
electroweak symmetry breaking), rather than $M_{GUT}$ or $M_{Pl}$. It also ameliorates the mild fine tuning that now must be imposed on the MSSM due to the LEP II constraints on the Higgs mass. 

The structure of the NMSSM can be used to motivate an extension of the GMSB messenger sector to include such an extra
Higgs singlet field, $N$ \cite{han2}.  The superpotential for this sector is then given by:
\begin{equation}
W = \xi_S S \bar{\Phi} \Phi + \xi_N N   \bar{\Phi} \Phi - \frac{\eta_S}{2} S^2 N - \frac{\eta_N}{2}N^2 S + \lambda_S S H_u H_d + 
\lambda_N N H_u H_d -\frac{k}{3} N^3 .
\label{superpot}
\end{equation}
Here $S$ is a Standard Model gauge singlet spurion which parameterizes
the supersymmetry breaking sector, with expectation values for the
scalar and supersymmetry-breaking $F$ components, $\langle S \rangle
\equiv M \neq 0$ and $\langle F_S \rangle \equiv F \neq 0$.  The
messenger fields $\Phi$ and $\bar{\Phi}$ are in complete ${\bf 5}$ and
${\bar{\bf 5}}$ representations of
SU(5)$\supset$SU(3)$\times$SU(2)$\times$U(1), so contain SU(3) triplet
states with the gauge quantum numbers of the right handed down quarks and
SU(2)-doublet states with the gauge quantum numbers of the left handed lepton
doublet.  The Higgs doublets $H_u$ and $H_d$ couple to up and down
type fermions, respectively.  The parameters $\xi_S, \xi_N, \eta_N,
\lambda_N$ and $k$ are expected to be $\mathcal{O}(1)$ while
$\eta_S$ and $\lambda_S$ have to to be considerably smaller ($\sim$10$^{-3}$) to give a consistent phenomenology for the $\mu$ and $B_{\mu}$
parameters~\cite{han2}.  At the tree level, the fermion components of
the $\Phi +\bar{\Phi}$ fields have, as a result of $\langle S
\rangle$, a common Dirac mass, $M$, while because of susy-breaking the
scalar components split into two groups with mass-squared $M^2\pm F$.
At the one-loop level, the masses of the lighter group of scalars are
further split, and the lightest of these scalars is the dark matter candidate.
A computation shows that the color-charged scalars are significantly
raised in mass, while the splitting between the lightest electrically
charged, $\phi^+$, and neutral, $\phi^0$, messenger is~\cite{dgp}
\begin{equation}
m_{\phi^+} - m_{\phi^0} \simeq \frac{\alpha M_Z^2}{4 \pi (M^2-F)^{1/2}}
\left( 4 \log\left[\frac{F}{M^2-F}\right]
-\log\left[\frac{M^2+F}{M^2-F}\right] +
\frac{2F}{M^2-F}\log\left[\frac{2F}{M^2+F}\right] -4 \right) .
\label{split}
\end{equation}
A physically acceptable model of messenger dark matter requires that
this splitting be positive, so that the neutral scalar is the lightest
state, and large enough such that charged state $\phi^+$ decays 
via $\phi^+\rightarrow \phi^0 e^+ \nu$ to avoid the presence of exotic quasi-stable charged particles.\footnote{In Ref.\cite{dgp} it was argued that
the decay had to occur before big-bang-nucleosynthesis, imposing the
stringent constraint $m_{\phi^+} - m_{\phi^0}\ge 5$~MeV.  This is too
restrictive, however, as for messenger masses in the 20 TeV range under
consideration the total energy deposited from such decays is
well below the limits set by BBN photodisociation.}
This requires $m_{\phi^+} - m_{\phi^0}\gsim 0.6$~MeV, which for
$M=50$~TeV imposes $F/M^2 \gsim 0.93$, leading to $m_{\phi^0}\lsim 13$~TeV,
while for $M=150$~TeV, $F/M^2 \gsim 0.96$ is necessary, leading
to $m_{\phi^0}\lsim 30$~TeV. Thus we see that the physically relevant
range of candidate dark matter messenger masses is $\sim 13 - 30$~TeV.

The potential that results from the minimizing of the superpotential
Eqn.~\ref{superpot} includes the terms:
\begin{equation}
V = (4 \xi_S \xi_N-2 \xi_N \eta_N) S N \bar{\Phi} \Phi + 2 \eta_N k S N^3 - 2 \eta_N \lambda_N N S H_u H_d + ...
\end{equation}
These three terms represent the vertices $N$-$\bar{\Phi}$-$\Phi$, $N$-$N$-$N$ and $N$-$H_u$-$H_d$, respectively, each with a 
coupling enhanced by the vacuum expectation value of $S$, $\langle S \rangle\sim 100$ TeV. It is because of this large vacuum 
expectation value that other terms in the potential which do not contain a factor of $S$ can be safely neglected in our calculation. 
Furthermore, we can neglect terms that include the small parameters
$\eta_S$ and $\lambda_S$.

These couplings lead to the following diagrams for messenger annihilation:

\vspace{-0.3cm}

\begin{feynartspicture}(222,204)(3,4.3)
\FADiagram{ }
\FAProp(0.,20.)(15.,10.)(0.,){/Straight}{+1}
\FALabel(-5.,18.0)[b]{$\Phi$}
\FAProp(15.,-5.0)(15.,10.0)(0.,){/Straight}{-1}
\FAProp( 0.,-15.0)(15.,-5.0)(0.,){/Straight}{-1}
\FALabel(-5.,-18.0)[b]{$\bar{\Phi}$}
\FALabel(10.,0.0)[b]{$\Phi$}
\FAProp(15.,10.)(30.,20.)(0.,){/ScalarDash}{0}
\FAProp(15.,-5.)(30.,-15.0)(0.,){/ScalarDash}{0}
\FALabel(35.,21.5)[t]{$N$}
\FALabel(35.,-15.)[t]{$N$}
\end{feynartspicture}

\vspace{-7.3cm}

\begin{feynartspicture}(222,204)(3,4.3)
\FADiagram{ }

\FAProp(50.,14.)(65.,3.)(0.,){/Straight}{+1}
\FALabel(55.,14.0)[b]{$\Phi$}

\FAProp(50.,-7.0)(65.,3.0)(0.,){/Straight}{-1}
\FALabel(55.,-10.0)[b]{$\bar{\Phi}$}

\FAProp(65.,3.)(85.,3.0)(0.,){/ScalarDash}{0}
\FALabel(75.,10.)[t]{$N$}

\FAProp(100.,14.)(85.,3.0)(0.,){/ScalarDash}{0}
\FALabel(95.,17.)[t]{$N$}

\FAProp(100.,-7.)(85.,3.0)(0.,){/ScalarDash}{0}
\FALabel(95.,-7.)[t]{$N$}
\end{feynartspicture}

\vspace{-7.3cm}

\begin{feynartspicture}(222,204)(3,4.3)
\FADiagram{ }

\FAProp(115.,14.)(130.,3.)(0.,){/Straight}{+1}
\FALabel(120.,14.0)[b]{$\Phi$}

\FAProp(115.,-7.0)(130.,3.0)(0.,){/Straight}{-1}
\FALabel(120.,-10.0)[b]{$\bar{\Phi}$}

\FAProp(130.,3.)(150.,3.0)(0.,){/ScalarDash}{0}
\FALabel(140.,10.)[t]{$N$}

\FAProp(165.,14.)(150.,3.0)(0.,){/ScalarDash}{0}
\FALabel(160.,18.)[t]{$H_u$}

\FAProp(165.,-7.)(150.,3.0)(0.,){/ScalarDash}{0}
\FALabel(160.,-7.)[t]{$H_d$}
\end{feynartspicture}

\vspace{-3.0cm}

The thermally averaged annihilation cross sections corresponding to these diagrams are given by:
\begin{eqnarray}
\langle\sigma v\rangle_{\phi \phi \rightarrow NN} \simeq \frac{\langle S\rangle^4}{256 \pi m_{\phi^0}^6} \bigg(\xi_N^4 (2 \xi_S 
-\eta_N)^4+ 
 \frac{1}{4}\bigg[1-\frac{1}{\sqrt{2}}\bigg]\eta_N^2 k^2  (2 \xi_S -\eta_N)^2 - \frac{1}{2}\xi_N^3 (2 \xi_S -\eta_N)^3 \eta_N k   
\bigg) \nonumber \\
 + \frac{\langle S\rangle^4}{256 \pi m_{\phi^0}^6} \bigg(-\frac{9}{2}\xi_N^4 (2 \xi_S -\eta_N)^4 - 
 \frac{3}{2}\bigg[1-\frac{1}{\sqrt{2}}\bigg]\eta_N^2 k^2  (2 \xi_S -\eta_N)^2 + \frac{11}{4}\xi_N^3 (2 \xi_S -\eta_N)^3 \eta_N k   
\bigg)\frac{T_{FO}}{m_{\phi^0}}, \,\,\,
\label{NN}
\end{eqnarray}
and
\begin{eqnarray}
\langle\sigma v\rangle_{\phi \phi \rightarrow H_u H_d} \simeq \frac{\langle S\rangle^4 \eta_N^2 \lambda_N^2 \xi_N^2 (2 \xi_S 
-\eta_N)^2  }{512 \pi m_{\phi^0}^6} 
 -   \frac{3 \langle S\rangle^4 \eta_N^2 \lambda_N^2 \xi_N^2 (2 \xi_S -\eta_N)^2  }{256 \pi m_{\phi^0}^6}    
\frac{T_{FO}}{m_{\phi^0}}, \,\,\,
\label{HH}
\end{eqnarray}
where $T_{FO}$ is the temperature at which $m_{\phi^0}$ ``freezes out'' of thermal equalibrium, given by
\begin{equation}
\bigg(\frac{T_{FO}}{m_{\phi^0}}\bigg)^{-1} \simeq \ln\bigg(c(c+2)\sqrt{\frac{45}{8}} \frac{g \, m_{\phi^0} \, M_{Pl} \, 
\langle\sigma v\rangle}{2 \pi^3 \, g^{\star}}\, \sqrt{\frac{T_{FO}}{m_{\phi^0}}}\,\, \bigg),
\end{equation}
where $M_{Pl}$ is the Planck mass, $c$ is an  $\mathcal{O}(1)$ constant and $g^{\star}$ is the number of degrees of freedom below the freeze out temperature ($\simeq 
232$ for the particle content of the NMSSM). When solved by iteration, values of approximately $T_{FO} \simeq m_{\phi^0}/20$ are 
found. With this in mind, we can see by inspecting Eqns.~\ref{NN} and~\ref{HH} that the temperature independent portion of the 
annihilation cross section (the first term in each expression) dominates at the freeze-out temperature. After freeze-out, the relic 
density of $\phi^0$ which remains today is given by
\begin{equation}
\Omega_{\phi^0} h^2 \simeq 5.6 \times 10^{-12} \, \frac{m_{\phi^0}}{T_{FO}} \bigg( \frac{\rm{GeV}^{-2}} {\langle\sigma 
v\rangle}\bigg) \simeq 0.1 \times  \bigg( \frac{1.3 \times 10^{-26} \, \rm{cm}^3/\rm{s}}{\langle\sigma v\rangle} \bigg).
\end{equation}
Thus we conclude that to obtain the cosmologically measured density of dark matter ($\Omega_{DM} h^2 \simeq 0.1$), we must require 
$\langle\sigma v\rangle \simeq 1.3 \times 10^{-26} \, \rm{cm}^3/\rm{s} \,\,$ ($1.1 \times 10^{-9} \, \rm{GeV}^{-2}$). Assuming 
$\xi_S, \xi_N, \eta_N, \lambda_N$ and $k$ are order one constants, as they are expected to be, and $\langle S\rangle\sim$ 100 TeV, 
we can see from Eqns.~\ref{NN} and~\ref{HH} that the required cross section is consistent with $m_{\phi^0} \sim 10 \, \rm{TeV}$. In 
a parameter space scan of this model, the authors of Ref.~\cite{han2} have found masses in the range of approximately 7 to 24 TeV 
consistent with $\Omega_{\phi^0} h^2 \simeq 0.1$. This range is somewhat artificial, however, depending on the range over which parameters were allowed to vary.  

\section{Multi-TeV Gamma Rays From Messenger Dark Matter}

In a previous study of the HESS galactic center source \cite{horns}, it was found that the a WIMP with a mass in the range of 12 to 
29 TeV could reproduce the observed spectrum, with the best fit occurring for 19 TeV. This conclusion depends, however, on the 
assumptions made regarding the annihilation products of the WIMP. The results of Ref.~\cite{horns} are therefore not completely 
general. In the case we are studying here, the dark matter candidate does not annihilate directly to quark or gauge boson pairs, but 
rather to pairs of Higgs bosons. Although the dominant decays of these Higgs bosons depend on the parameters chosen in this sector, 
in most models, neutral Higgses decay to either heavy quarks or gauge bosons. Therefore, the gamma-ray spectrum produced in the 
annihilations of $\phi^0$'s resembles the spectrum found in Ref.~\cite{horns}, but for a WIMP with half the mass.

\begin{figure}[t]
\centering\leavevmode
\mbox{
\includegraphics[width=2.3in,angle=90]{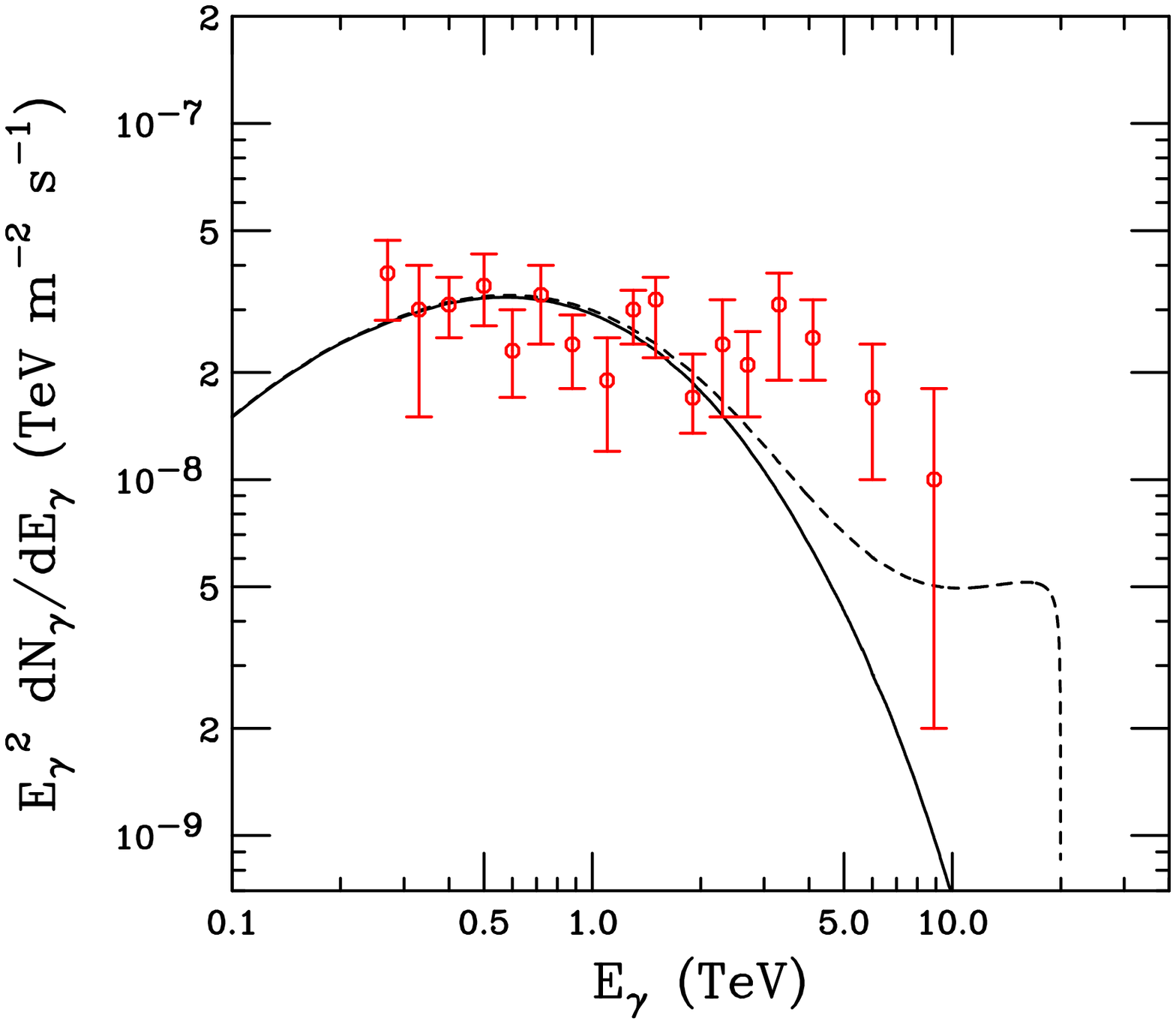}
\hfill
\includegraphics[width=2.3in,angle=90]{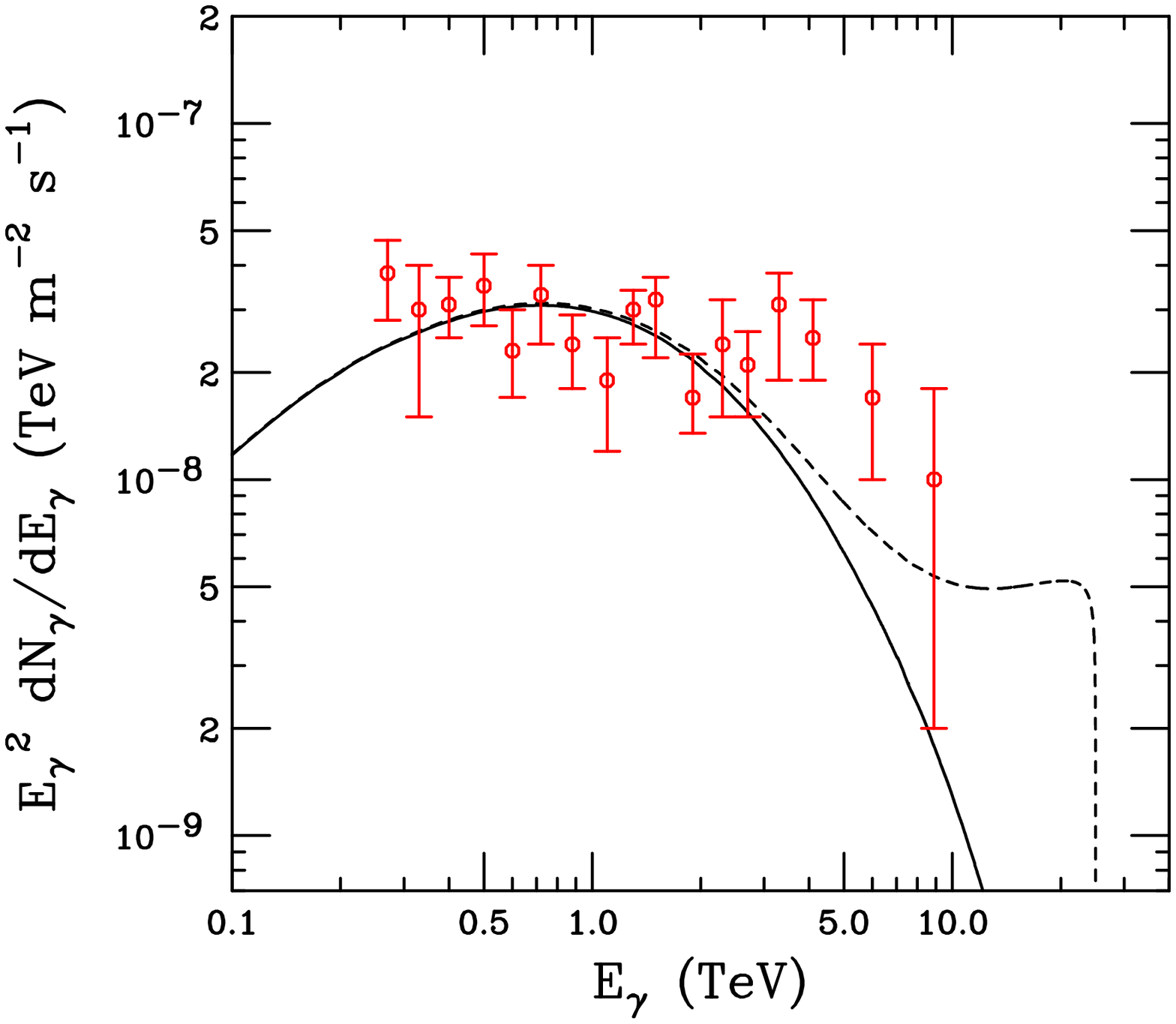}}
\caption{The gamma-ray spectrum predicted from the annihilation of messenger dark matter particles ($\phi^0$'s) compared to the HESS 
data \cite{hess}. The spectra shown represent the spectra for annihilations into 100\% neutral Higgses (solid lines) and 50\% 
neutral and 50\% charged Higgses (dashed lines). The left and right frames are for 20 and 25 TeV dark matter states, respectively. 
The results for normalized to best fit the HESS data. See the text for more information.}
\label{spec}
\end{figure}

$\phi^0$ annihilations can also produce pairs of charged Higgs bosons. Although the gamma-rays produced in the decays of charged Higgses 
contribute minimally to the overall spectrum, annihilation diagrams with $H^+ H^- \gamma$ final states can contribute substantially 
at high energies. Such final states are guaranteed to be present, with the fraction of charged final states that include a 
radiated photon given by
\begin{equation}
\frac{dN_{\gamma}}{dx} = \frac{d(\sigma_{H^+ H^- \gamma} v)/dx}{\sigma_{H^+ H^-} v} \simeq \frac{\alpha}{\pi} \frac{(x^2-2x+2)}{x} 
\ln \bigg[\frac{m_{\phi^0}^2}{m_{H^{\pm}}^2}(1-x) \bigg],
\end{equation}
where $x = E_{\gamma}/m_{\phi^0}$. As we are considering $m_{\phi^0} \sim 20$ TeV and $m_{H^{\pm}}$ of a few hundred GeV, this 
contribution can be significant \footnote{Final state photons also play an important role for annihilating Kaluza-Klein dark matter 
which annihilate primarily to charge lepton pairs. See Ref.~\cite{bergstromkk}}. Of course this depends on the fraction of 
$\phi^0$ annihilations that go to charged Higgs pairs. From Eqns.~\ref{NN} and \ref{HH}, we can see that annihilations to either 
$NN$ or $H_u H_d$ can dominate, for appropriate choices of parameters. For annihilations to $H_u H_d$, approximately half of which 
produce $H^+ H^-$ final states, we can expect an important contribution from final state photons. Annihilations to $NN$ do not 
produce charged Higgses, as they mix only into neutral states. In figure~\ref{spec}, we show the spectrum of gamma-rays predicted 
for annihilations of $\phi^0$'s without a charged Higgs contribution (solid) and with 50$\%$ of annihilations producing a charged 
Higgs pair (dashed).  We have used $m_{h}=120$ GeV and $m_{A} \approx  m_{H} \approx  m_{H^{\pm}} \approx 300$ GeV, although our 
results depend very weakly on these choices. The spectra were generated using PYTHIA \cite{pythia} as implemented in the DarkSusy 
package \cite{darksusy}.

The gamma-ray spectra predicted from messenger dark matter annihilations can provide good fits to the HESS data. For 
the spectra shown in figure~\ref{spec}, with $m_{\phi^0} = 25$ TeV, we find fits with $\chi^2$ = 1.44 (solid line) and 1.20 (dashed 
line) per degree of freedom. For a somewhat lighter WIMP with $m_{\phi^0} = 20$ TeV, the quality drops to $\chi^2$ = 1.75 (solid 
line) and 1.36 (dashed line) per degree of freedom. For a 30 TeV WIMP, the $\chi^2$'s improve to 1.35 and 1.16 per degree of freedom with and without a charged Higgs component, respectively. It is clear that the presence of final state photons from charged Higgs diagrams improves the fit to the HESS data considerably.

In all cases, the overall normalization of the spectrum was selected to provide the 
best possible fit to the HESS data.

To normalize the gamma-ray spectrum from $\phi^0$ annihilations, the distribution of dark matter near the galactic center must be 
considered. Unfortunately, the central 
region of the galaxy is dominated by baryons, so observations of rotation curves reveal little about the distribution of dark matter in this region. 
Furthermore, simulations have limited resolution and have difficulties predicting the structure of dark matter halos below the 
kiloparsec scale. 

With these reservations in mind, we will discuss some possibilities for the dark matter distribution in the central region of our 
galaxy. Perhaps the most popular class of distributions are those found by groups using N-body simulations. These include the well 
known Navarro Frenk and White (NFW) \cite{Navarro:1995iw} and Moore {\it et al.} \cite{Moore:1997sg} profiles. These distributions 
display a dark matter cusp in the central kiloparsecs of the galaxy with the behavior $\rho \propto 1/r^{\gamma}$ with $\gamma 
\simeq 1$ and 1.5 for the NFW and Moore {\it et al.} profiles, respectively. These N-body simulations do not include the effects of 
baryons on the dark matter distribution, however. In particular, it has recently been argued that the cooling of baryons in the 
inner kiloparsecs of our galaxy will compress the dark matter distribution, resulting in considerably higher densities of dark 
matter~\cite{adiabatic}. The conclusions drawn regarding this process, called adiabatic compression, are still somewhat 
controversial, however. Thirdly, the presence of a $2.6 \times 10^6$ solar mass black hole present at the dynamical center of our 
galaxy has lead some to argue that the accretion of dark matter onto this body could generate a density spike with 
$\rho \propto 1/r^{2.4}$ \cite{spike1,spike2}. Others have challenged this conclusion, however \cite{nospike}. 

The gamma-ray flux produced in the annihilations of dark matter particles near the galactic center is given by
\begin{equation}
\Phi_{\gamma}(\psi, E_{\gamma}) = \langle\sigma v\rangle_{v \rightarrow 0} \frac{dN_{\gamma}}{dE_{\gamma}}  \frac{1}{4 \pi 
m_{\phi^0}^2} \int_{los} ds \rho^2(r),
\end{equation}
where $\psi$ is the angle from the galactic center, $\langle\sigma v\rangle_{v \rightarrow 0}$ is the WIMP's annihilation cross 
section in the non-relativistic limit, $dN_{\gamma} / dE_{\gamma}$ is the differential spectrum of gamma-rays produced in each 
annihilation and $\rho(r)$ is the dark matter density at a distance, $r$, from the galactic center.  The integral is performed over 
the line-of-sight of the observation.

We can break this expression into two components, one of which depends only on the properties of the dark matter particle and 
another which depends only on the dark matter distribution. The second of these factors is given by
\begin{equation}
J(\psi) = \frac{1}{8.5 \, \rm{kpc}} \bigg(\frac{1}{0.3 \, \rm{GeV/cm}^3}\bigg)^2  \int_{los} ds \rho^2(r).
\end{equation}
We can then write
\begin{equation}
\Phi_{\gamma}(\psi, E_{\gamma}) \cong 6 \times 10^{-11} \, \frac{dN_{\gamma}}{dE_{\gamma}} \bigg(\frac{\langle\sigma v\rangle_{v 
\rightarrow 0}}{1.3 \times 10^{-26}\rm{cm}^3/\rm{s}}\bigg) \bigg(\frac{20 \, \rm{TeV}}{m_{\phi^0}} \bigg)^2 \,  \overline{J(\Delta 
\Omega)} \Delta \Omega \,\,\rm{m}^{-2} \,\rm{s}^{-1},
\end{equation}
where $\overline{J(\Delta \Omega)}$ is the average of $J(\psi)$ over the solid angle $\Delta \Omega$ (centered on $\psi=0$). 
Considering the range of masses and annihilation modes discussed in this letter, and a solid angle consistent with the upper limit 
reported by HESS ($\Delta \Omega \simeq 3 \times 10^{-6}$ steradians), we estimate approximately $3 \times 10^8$ to be the value of 
$\overline{J(3 \times 10^{-6} \rm{sr})}$ required to generate the observed flux. The profiles favored by N-body simulations predict 
considerably smaller values, with $\overline{J(3 \times 10^{-6} \rm{sr})} \sim 3 \times 10^4$ and $4 \times 10^7$ for the NFW and 
Moore {\it et al.} profiles, respectively. These could potentially be increased by a factor of 10 to 100 due to adiabatic 
compression, however, reaching the desired range. The presence of a dark matter density spike from adiabatic accretion onto the 
central supermassive black hole could also potentially generate such large values.

\section{Conclusions}

The origin of the recently observed multi-TeV gamma-rays from the galactic center region is not presently understood. Although 
astrophysical possibilities have been proposed, annihilating dark matter is another possible explanation. The most popular dark matter 
candidates, such as the lightest Neutralino in models of Supersymmetry, are far too light too account for the very high-energy 
spectrum observed.

In this letter, we have considered the possibility that stable particles from the messenger sector of Gauge Mediated Supersymmetry 
Breaking (GMSB) models are the major component of the dark matter of the universe and that the multi-TeV gamma-ray spectrum observed 
by HESS is the product of their annihilations. We find that 20 to 30 TeV messenger dark matter matter particles can reproduce the 
observed spectrum. The overall flux can be produced only if the density of dark matter is very large in the inner several parsecs of 
the galaxy, perhaps as the result of adibatic compression of the halo due to baryon cooling or the adibatic accretion of dark matter 
onto the central supermassive black hole.

\vspace{0.5cm}

{\it Acknowledgments}: We would like to thank Joe Silk for insightful discussions. DH is supported by the Leverhulme Trust. \vskip 
-0.5cm

\end{document}